\documentclass[aps,prb,twocolumn,superscriptaddress,showpacs,floatfix,preprintnumbers,amsmath,amssymb]{revtex4}
\usepackage{graphicx,epsfig,amsmath,amssymb,natbib}
\newcommand\cco{$\rm{Ca_3Co_2O_6}$}
\newcommand\Tn{$T_{\mathrm{N}}$} 
\newcommand\AF{antiferromagnet}

\begin{document}
\title{Spin Correlations in Ca$_3$Co$_2$O$_6$: A Polarised-Neutron Diffraction and Monte Carlo Study}
\author{Joseph~A.~M.~Paddison} \affiliation{Department of Chemistry, University of Oxford, Inorganic Chemistry Laboratory, South Parks Road, Oxford OX1 3QR, United Kingdom} \affiliation{ISIS Facility, Rutherford Appleton Laboratory, Chilton, Didcot, Oxfordshire OX11 0QX, U.K.}
\author{Stefano~Agrestini} \affiliation{Max Planck Institute for Chemical Physics of Solids, N\"othnitzerstr. 40, 01187 Dresden, Germany}
\author{Martin~R.~Lees} \affiliation{Department of Physics, University of Warwick, Coventry, CV4 7AL, United Kingdom}
\author{Catherine~L.~Fleck} \affiliation{Department of Physics, University of Warwick, Coventry, CV4 7AL, United Kingdom}
\author{Pascale~P.~Deen} \affiliation{Institut Laue-Langevin, 6 rue Jules Horowitz, 38042 Grenoble, France} \affiliation{European Spallation Source, ESS AB, SE-22100 Lund, Sweden}
\author{Andrew~L.~Goodwin} \affiliation{Department of Chemistry, University of Oxford, Inorganic Chemistry Laboratory, South Parks Road, Oxford OX1 3QR, United Kingdom}
\author{J.~Ross~Stewart} \affiliation{ISIS Facility, Rutherford Appleton Laboratory, Chilton, Didcot, Oxfordshire OX11 0QX, U.K.}
\author{Oleg~A.~Petrenko} \email[]{O.Petrenko@warwick.ac.uk} \affiliation{Department of Physics, University of Warwick, Coventry, CV4 7AL, United Kingdom}  
\date{\today}

\begin{abstract}
We present polarised-neutron diffraction measurements of the Ising-like spin-chain compound \cco\ above and below the magnetic ordering temperature \Tn.
Below \Tn, a clear evolution from a single-phase spin-density wave (SDW) structure to a mixture of SDW and commensurate \AF\ (CAFM) structures is observed on cooling.
For a rapidly-cooled sample, the majority phase at low temperature is the SDW, while if the cooling is performed sufficiently slowly, then the SDW and the CAFM structure coexist between 1.5 and 10~K.
Above \Tn, we use Monte Carlo methods to analyse the magnetic diffuse scattering data.
We show that both intra- and inter-chain correlations persist above \Tn, but are essentially decoupled.
Intra-chain correlations resemble the ferromagnetic Ising model, while inter-chain correlations resemble the frustrated triangular-lattice \AF.
Using previously-published bulk property measurements and our neutron diffraction data, we obtain values of the ferromagnetic and \AF ic exchange interactions and the single-ion anisotropy.
\end{abstract}
\maketitle
\section{Introduction}
The spin-chain compound \cco\ offers a rare opportunity to investigate the interplay of frustration and low dimensionality in a triangular lattice of Ising chains.
Some of the most interesting observations of \cco\ have been made in an applied magnetic field, where the appearance of magnetisation plateaux\cite{Hardy_2004} at low temperature has been linked to a quantum tunnelling mechanism.\cite{Maignan_2004}
However, even in zero field \cco\ demonstrates very complex and unexpected behaviour, including an unusual order--order transition, an ultra-slow magnetic relaxation, and a coexistence of several magnetic phases.\cite{Agrestini_2011}
A substantial and prolonged theoretical interest in this compound~\cite{Wu_2005,Kudasov_2006,Kudasov_2007,Yao_2006,Kamiya_2012} is thus not unexpected (see Ref.~\onlinecite{Kudasov_2012} for a recent review).

The crystal structure of \cco\ is hexagonal (space group $R\bar{3}c$) and consists of chains made up of alternating face-sharing octahedral (Co$^{3+}_{\rm{I}};S=0$) and trigonal prismatic (Co$^{3+}_{\rm{II}};S=2$) CoO$_6$ polyhedra.\cite{Fjellvag_1996,Takubo_2005}
The chains are directed along the $c$ axis and are arranged on a triangular lattice in the $ab$ plane. The trigonal crystal field and spin-orbit coupling generate an Ising-like magnetic anisotropy at the Co$^{3+}_{\rm{II}}$ site, with the easy axis parallel to the chains.\cite{Kageyama_1997,Wu_2005}
Spins are coupled by ferromagnetic (FM) interactions within the chains, while much weaker \AF ic (AFM) interactions couple adjacent chains along helical paths.\cite{Fresard_2004,Chapon_2009}
Below $T_{\rm{N}} = 25$\,K a magnetic order is stabilised in the form of a longitudinal amplitude-modulated spin-density wave (SDW) propagating along the $c$ axis with a periodicity of about 1000~\AA.
There is a phase shift of $120^\circ$ in the modulation between adjacent chains.\cite{Agrestini_2008,Agrestini_2008a}
Recent powder neutron diffraction results\cite{Agrestini_2011} revealed an order--order transition from the SDW structure to a commensurate \AF ic (CAFM) phase.
This transition occurs over a time-scale of several hours and is never complete.

Here, we report polarised-neutron scattering measurements on a polycrystalline sample of \cco\ above and below \Tn, supported by Monte Carlo simulations of the magnetic diffuse scattering and bulk properties above \Tn.
Our paper is structured as follows. Details of our experiments and analysis methods are given in Section~\ref{sec:technical}.
In Section~\ref{sub:low_temp}, we address the effect of temperature and sample cooling rate on the coexistence of magnetic phases below \Tn.
In Section~\ref{sub:high_temp}, we use Monte Carlo methods to extract the spin correlation functions parallel and perpendicular to the chains above \Tn.
Finally, in Section~\ref{sec:bulk} we develop a model of the magnetic interactions which is compatible both with our neutron data and with previously-reported measurements of bulk properties.\cite{Maignan_2004,Hardy_2003,Hardy_2003a}

\section{Experimental and Technical Details}\label{sec:technical}
A polycrystalline sample of \cco\ of mass 3.6~g was synthesised via a solid-state method.\cite{Kageyama_1997,Kageyama_1997a,Maignan_2000}
The magnetic properties of the sample were checked by magnetisation measurements and agree with those reported previously.\cite{Kageyama_1997,Kageyama_1997a,Maignan_2000,Hardy_2003a,Hardy_2004a}

We performed polarised-neutron diffraction experiments using the D7 instrument at the ILL in Grenoble, France.
D7 is a  cold-neutron diffuse scattering spectrometer equipped with $xyz$ polarisation analysis,\cite{Stewart_2009} which uses 132 $^3$He detectors to cover a scattering range of about $140^\circ$.
We used neutrons monochromated to a wavelength of 4.8 and  3.1~\AA, allowing the scattering to be measured in the range $0.34<q<2.48$~\AA$^{-1}$ and $0.53<q<3.46$~\AA$^{-1}$ respectively.
In order to avoid possible complications from the influence of the out-of-plane components to forward scattering\cite{Ehlers_2013} the data for a scattering angle less than $20^\circ$ have been ignored in the refinements.

Standard data analysis techniques (which included detector efficiency normalisation from vanadium standards and polarisation efficiency calculations from an amorphous silica standard) were used to calculate the magnetic, nuclear and spin-incoherent scattering components from the initial data.
The diffraction data obtained below \Tn\ were refined using the {\small FULLPROF} program.\cite{Carvajal_1993}
D7 is also equipped with a Fermi chopper, which permits inelastic scattering measurements with a resolution of 3\% of the incident energy, thereby giving the ability to differentiate between truly elastic scattering and quasi-elastic or inelastic contributions.
All attempts to detect an inelastic signal at any temperature were unsuccessful; therefore within D7's resolution of about 0.15~meV the signal should be presumed to be totally elastic. 

We refined the magnetic diffuse scattering patterns obtained above \Tn\ using the Spinvert program,\cite{Paddison_2013} which implements a reverse Monte Carlo (RMC) algorithm.\cite{McGreevy_1988, Paddison_2012}
In the RMC refinements, a supercell of the crystallographic unit cell is first generated, a classical Ising spin with random orientation (up/down) is assigned to each Co$^{3+}_{\rm{II}}$ site, and the sum of squared residuals is calculated:
\begin{equation}
\chi^2=W\sum_q\left[\frac{sI_{\mathrm{calc}}(q)-I_{\mathrm{expt}}(q)}{\sigma(q)}\right]^2,
\label{eq:chi sq}
\end{equation}
in which $I(q)$ denotes a powder-averaged magnetic scattering intensity, subscripts ``$\mathrm{calc}$" and ``$\mathrm{expt}$" denote calculated and experimental values, $\sigma(q)$ is an experimental uncertainty, $W$ is an empirical weighting factor, and $s$ is a refined overall scale factor.
A randomly-chosen spin is then flipped, the change in $\chi^2$ is calculated, and the proposed spin-flip is accepted or rejected according to the Metropolis algorithm.
This process is repeated until no further reduction in $\chi^2$ is observed.
The scattering pattern is calculated from the spin configuration using the general expression of Ref.~\onlinecite{Blech_1964} which takes magnetic anisotropy into account.
In our refinements we used periodic spin configurations of size $12\times 7\times 11$ orthorhombic \footnote[29]{The orthorhombic axes are related to the hexagonal axes by $\mathbf{a}_{\rm{O}}=\mathbf{a}_{\rm{H}}$, $\mathbf{b}_{\rm{O}}=\mathbf{a}_{\rm{H}}+2\mathbf{b}_{\rm{H}}$, $\mathbf{c}_{\rm{O}}=\mathbf{c}_{\rm{H}}$.} unit cells ($N=11088$ spins), which represented a compromise between maximising the simulation size and keeping the computer time required within reasonable limits.
Refinements were performed for 100 proposed flips per spin, and all calculated quantities were averaged over 10 independent spin configurations in order to minimise the statistical noise.

We also performed direct Monte Carlo (DMC) simulations using the Ising Hamiltonian,
\begin{equation}
H=-\frac{1}{2}\sum_{i,j}J_{ij}S^{z}_i S^{z}_j,
\label{ising_hamiltonian}
\end{equation}
and the anisotropic Heisenberg Hamiltonian,
\begin{equation}
H=-\frac{1}{2}\sum_{i,j}J_{ij}\mathbf{S}_i\cdot\mathbf{S}_j-D\sum_i {(S_i^z)}^2,
\label{heisenberg_hamiltonian}
\end{equation}
for various sets of magnetic interactions relevant to \cco.
Here, $\mathbf{S}_i$ is a classical spin vector with $z$-component $S_i^z$ and length $|\mathbf{S}|$, $J_{ij}$ is a general interaction between pairs of spins $i,j$, $D$ is a single-ion anisotropy term, and the factor of $1/2$ corrects for the double-counting of pairwise interactions.
In Section~\ref{sub:high_temp}, we consider specifically interactions between nearest-neighbour and third-neighbour spins, which we label $J_1$ and $J_3$ respectively.

Direct Monte Carlo simulations were performed using a simulated annealing algorithm in which a periodic spin configuration was initialised with random spin orientations and slowly cooled.
The initial temperature was $T=15J_1 |\mathbf{S}|^2$ for the Ising model and $T=60J_1|\mathbf{S}|^2$ for the anisotropic Heisenberg model, and ratio of adjacent temperatures was equal to 0.96.
At each temperature, $10^4$ moves per spin were proposed for equilibration, followed by at least $10^5$ proposed moves for calculations of the bulk properties.
For the Ising model, a proposed spin move involved choosing a spin at random and flipping its orientation.
For the anisotropic Heisenberg model, two kinds of move were alternated: randomly choosing a new spin orientation on the surface of a sphere, and a simple spin flip ($\mathbf{S}_{i}\rightarrow-\mathbf{S}_{i}$), where the latter is used to allow the simulation to move rapidly between Ising-like states at low $T$.\cite{Hinzke_1999}
Each proposed spin move was accepted or rejected according to the Metropolis algorithm.
To simulate diffraction patterns, we used 5 independent spin configurations, each of size $18\times 10 \times 16$ orthorhombic unit cells ($N=34560$~spins); an approximately cubic supercell was used because calculating powder diffraction patterns involves spherically-averaging spin correlation functions in real space.\cite{Blech_1964}
For bulk-properties calculations, we used a spin configuration of size of $5\times 3 \times 64$ orthorhombic unit cells ($N=11520$~spins); these dimensions were chosen since the correlation length along $c$ is longer than in the $ab$ plane [Section~\ref{sub:high_temp}].
The magnetic susceptibility per spin, $\chi$, and heat capacity per spin, $C_{\rm{mag}}$, were calculated from the fluctuation-dissipation relations,
\begin{eqnarray}
\chi T & = & \frac{1}{N} \left(\left\langle M_z^2\right\rangle -\left\langle M_z\right\rangle ^2\right),\\
C_{\mathrm{mag}} & = & \frac{1}{NT^2} \left(\left\langle E^2\right\rangle -\left\langle E\right\rangle ^2\right),
\end{eqnarray}
where $M_z=\sum_iS_i^z/|\mathbf{S}|$ is the total magnetisation (normalised by spin length), $E=\sum_iE_i$ the total energy, and angle brackets denote the time average.
For comparison with experimental data, $\chi T$ is converted into units of $\mathrm{K\, m^3\, mol^{-1}}$ by multiplying by $N_{\mathrm{A}}\mu_0\mu_{\mathrm{eff}}^2k_{\mathrm{B}}^{-1}$, with the effective magnetic moment $\mu_{\mathrm{eff}}=g|\mathbf{S}|\mu_\mathrm{B}$, and $C_{\mathrm{mag}}$ is converted into $\mathrm{J\, K^{-1}\, mol^{-1}}$ by multiplying by $N_{\mathrm{A}}k_{\mathrm{B}}$. 

\section{Results and Discussion}
\subsection{Low-Temperature Data}{\label{sub:low_temp}}
\begin{figure}[tb]
\begin{center}
\includegraphics[width=0.9\columnwidth]{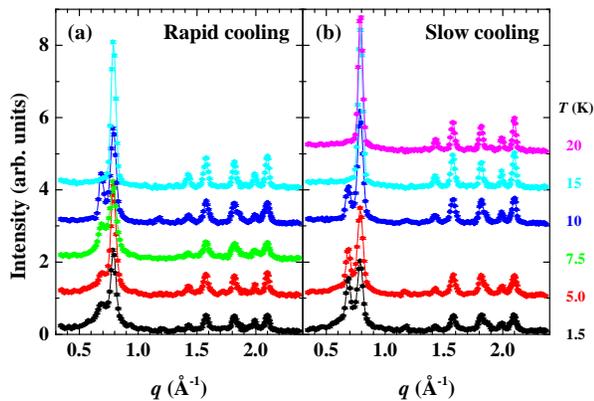}
\caption{\label{fig1_IQ}	Magnetic powder neutron diffraction intensity profiles of \cco\ measured with 4.8~\AA\ neutrons on warming after rapid cooling (a) and on slow cooling (b).
					The curves are offset for clarity.}
\end{center}
\end{figure}
The CAFM phase detected previously in \cco\ has been proven to be metastable in nature.\cite{Agrestini_2011}
Given this, we have experimented using two different protocols for cooling the sample, denoted {\it slow cooling} and {\it rapid cooling}.
For the slow cooling, the sample was initially cooled to 30\,K; the temperature was then reduced down to 5~K in steps of 5~K, and finally to the base temperature of 1.5~K.
Diffraction patterns were recorded at each temperature step with a data collection time of 4 to 5 hours; therefore the total cooling time to base temperature was more than 24~hours.
For the rapid cooling, the sample was quickly (within a few minutes) cooled from just above the ordering temperature ($\sim 25$~K) down to the base temperature of the cryostat and equilibrated for 15~minutes.
The measurement time at the base temperature was 4~hours.
The sample was then warmed up to 5, 7.5, 10 and 15~K with 4-hour long measurements at each temperature.
The neutron diffraction patterns recorded with 4.8~\AA\ neutrons following these two protocols are shown in Figs.~\ref{fig1_IQ}(a) and \ref{fig1_IQ}(b) respectively.

The most intense magnetic Bragg peaks corresponding to the SDW and CAFM phases appear at 0.79 and 0.69~\AA $^{-1}$ respectively.
The presence of these two magnetic phases can therefore be easily followed in Fig.~\ref{fig1_IQ}.

\begin{figure}[tb]
\begin{center}
\includegraphics[width=0.8\columnwidth]{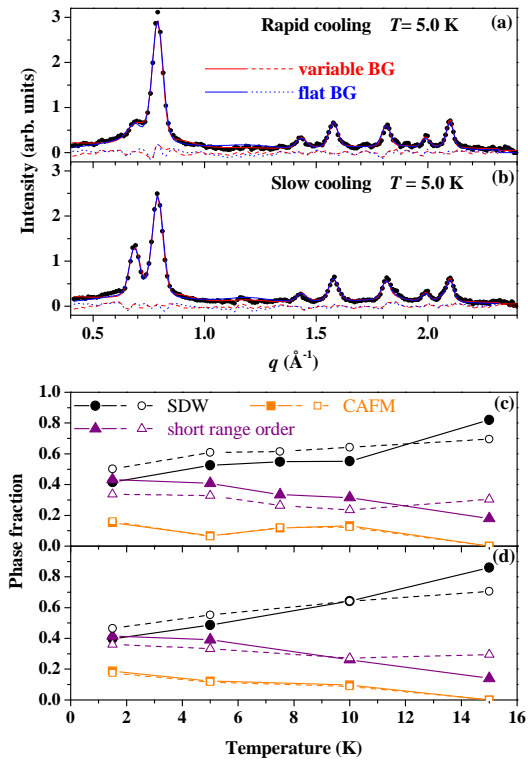}
\caption{\label{fig2_IvsQ_Fractions}	Top panels: Magnetic component of powder neutron-diffraction patterns of \cco\ collected at 5.0~K (a) on warming to this temperature after rapid cooling to 1.5~K and (b) on slow cooling to this temperature.
					Dots represent the experimental data, while the lines show the calculated patterns and difference curves.
					Bottom panels: Temperature dependence of the fractions of the SDW, CAFM and short-range order phases for rapid cooling (c) and slow cooling (d) protocols.
					Solid (open) symbols correspond to the the refinement which includes (excludes) a variable background.}
\end{center}
\end{figure}

The SDW phase is the majority phase at all temperatures.
The CAFM phase is present at temperatures below and including 10~K, but for the rapid cooling protocol it is barely visible at lower temperatures (1.5 and 5~K).
On the other hand for the slow cooling protocol intense and narrow magnetic reflections from the CAFM phase can be observed even at the lowest temperatures.
These results provide evidence that the extremely slow dynamics existing below 10~K hamper the development of the long range CAFM phase in the case of a fast cooling procedure.
Apart from the long-range SDW and CAFM phases, a short-range magnetic component is clearly present at $T<15$~K for both cooling protocols in agreement with the previous unpolarised neutron diffraction data.\cite{Agrestini_2011,Agrestini_2008}
In the refinement, it is rather difficult to distinguish this short-range component from a background signal that varies slightly as function of scattering angle.
Figs.~\ref{fig2_IvsQ_Fractions}(a) and \ref{fig2_IvsQ_Fractions}(b) illustrate this point by showing the refinement of the $T=5.0$~K data for rapid and slow cooling regimes using both flat and variable backgrounds.

The actual numbers for the phase fraction of the short-range component vary considerably depending on the presumed shape of the background.
The temperature dependence of the magnetic fractions are shown in Fig.~\ref{fig2_IvsQ_Fractions}(c) and Fig.~\ref{fig2_IvsQ_Fractions}(d) for both cooling regimes.

The slow cooling procedure results in the simultaneous presence of both magnetic long-range ordered phases at low temperature (from 1.5 to 10~K), while at higher temperatures (at 15 and 20~K) only the SDW phase is visible.
In contrast to the rapid cooling data, for the slow cooling procedure the fraction of the CAFM phase does not show a maximum around 10~K but monotonically increases, at the expense of the SDW phase, as the temperature is further reduced.
This result shows that the particular protocol used for cooling the sample strongly affects the evolution of the order-order transition between the SDW and CAFM phases.
This observation is an effect of the rapid increase of the characteristic time of the transition process between the SDW and CAFM phases as the temperature is decreased.\cite{Agrestini_2011}
The dependence on the cooling procedure shown by ac susceptibility measurements\cite{Hardy_2004a} is probably related to the particular dynamics of the long range magnetic order in \cco.

Due to the relatively low $q$-resolution of the D7 diffractometer, it was not possible to detect a small ($\sim 0.01$~\AA$^{-1}$)\cite{Agrestini_2008,Agrestini_2008a} incommensuration in the magnetic reflections associated with the long-period modulation of the SDW magnetic structure along the $c$ axis.

Additional data were collected using the slow cooling protocol with 3.1~\AA\ neutrons.
The higher-energy incident neutrons allowed for diffraction up to a higher maximum $q$, but the data collected essentially followed the same trend as in Fig.~\ref{fig1_IQ}(b) and are therefore not shown here.

\subsection{High-Temperature Data}{\label{sub:high_temp}}
\begin{figure}[tb]
\begin{center}
\includegraphics[width=0.85\columnwidth]{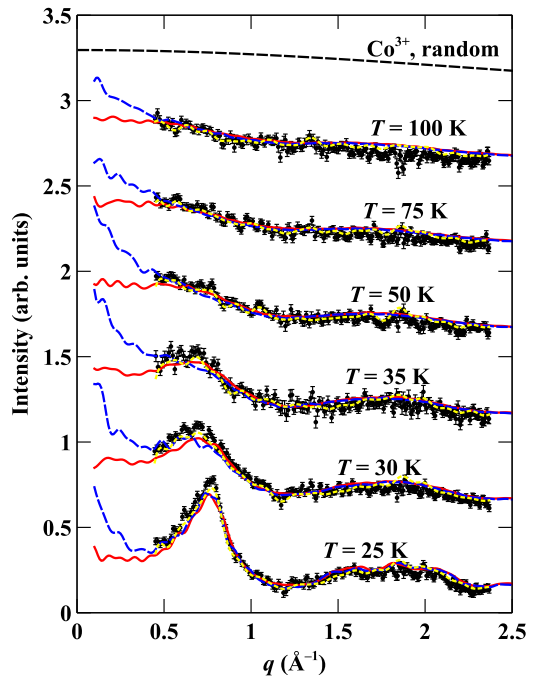}
\caption{\label{fig3_IQ}	Magnetic diffuse scattering patterns from \cco\ at six temperatures above \Tn.
					Filled black symbols represent experimental data. The dotted yellow lines show reverse Monte Carlo fits (see main text for details).
					The solid red lines are calculated using direct Monte Carlo simulations of the Ising model with exchange interactions $J_3/J_1=-0.10$ ($J_1|\mathbf{S}|^2=22.9$~K), and the dashed blue lines for $J_3/J_1=-0.02$ ($J_1|\mathbf{S}|^2=36.8$~K).
					The curves for different temperatures are consecutively offset by 0.5 units for clarity. The paramagnetic squared form factor for Co$^{3+}$ (adjusted by the overall intensity scale determined from RMC refinement) is shown as a dashed black line for comparison.}
\end{center}
\end{figure}
High-temperature magnetic diffuse scattering data, collected at six temperatures above \Tn\/, are shown in Fig.~\ref{fig3_IQ}.
No appreciable dependence of the scattering intensity on sample history was observed at any temperature above \Tn\/.
At $T=25$, 30 and 35~K, the magnetic diffuse scattering is dominated by a broad peak, which decreases in intensity and shifts from $q=0.8$ to 0.7~\AA$^{-1}$ with increasing temperature.
At $T\ge 50$~K the scattering patterns show no pronounced features in the $q$ range probed, but even at $T=100$~K there are small differences between the experimental data and the paramagnetic Co$^{3+}$ form factor which would be observed for entirely random spin orientations.
We use two approaches to analyse the high-temperature data.
First, we fit the data using reverse Monte Carlo refinement: this approach determines the paramagnetic correlations but does not model the magnetic interactions.
Second, we consider the extent to which our data are consistent with different sets of interactions in a simple magnetic Hamiltonian.
Throughout, we assume that only the Co$^{3+}_{\rm{II}}$ ($S=2$) sites are magnetic, with no magnetic moment present on the Co$^{3+}_{\rm{I}}$ ($S=0$) sites, and find that this assumption is entirely consistent with the data.

We performed RMC refinements of the high-temperature data using the Spinvert program.\cite{Paddison_2013}
Technical details of the refinements were given in Section~\ref{sec:technical}.
The most important assumption is that the spins behave as purely Ising variables.
Measurements of the magnetic susceptibility \cite{Hardy_2007,Cheng_2009} and our own analysis in Section~\ref{sec:bulk} suggest that this assumption is valid at least for $T\lesssim 50$~K, where the diffuse scattering shows the most pronounced features. We make the usual assumption that the magnetic form factor is given by the dipole (low-$q$) formula, $f(q)=j_{0}(q)+C_{2}j_{2}(q)$, where $j_{0}(q)$ and $j_{2}(q)$  are tabulated functions \cite{Brown_2004} and the factor $C_{2}=L_{z}/(2S_{z}+L_{z})$ accounts for the orbital contribution to the magnetic moment in \cco; we take $C_{2}=0.25$ as an average of the literature values.\cite{Hardy_2004,Maignan_2004,Wu_2005,Burnus_2006}
All fits are modified by an overall intensity scale factor in order to match the data.
The optimal value of this scale factor was obtained by fitting the $T = 25$~K data and then fixed at the $T = 25$~K values when fitting the higher-temperature datasets. The fits we obtain are shown in Fig.~\ref{fig3_IQ}, and represent excellent agreement with the data.

\begin{figure}[tb]
\begin{center}
\includegraphics[width=0.85\columnwidth]{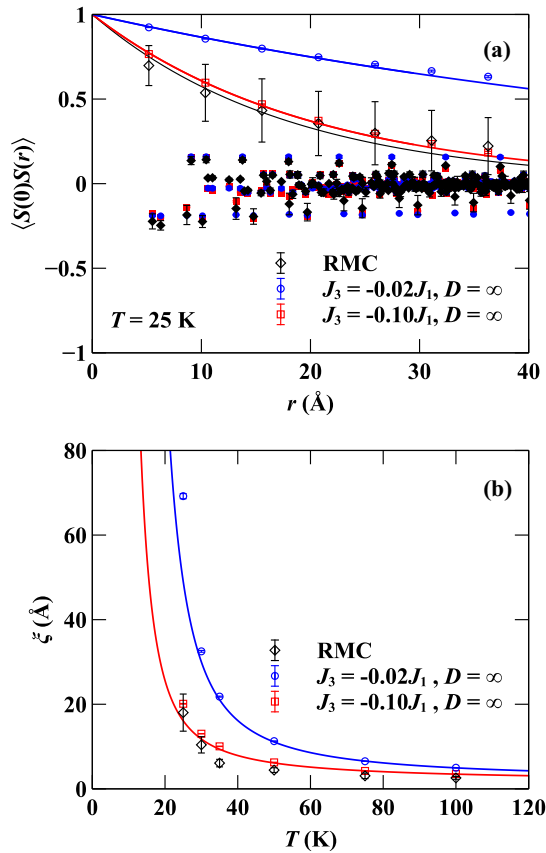}
\caption{\label{fig4_correlation}
	(a) Radial spin correlation function at $T=25$~K.
	Results from reverse Monte Carlo fits to the $T=25$~K data are shown in black.
	Values from the direct Monte Carlo simulations with $J_3/J_1=-0.10$ ($J_1|\mathbf{S}|^2=22.9$~K) and $J_3/J_1=-0.02$ ($J_1|\mathbf{S}|^2=36.8$~K) are coloured red and blue, respectively.
	Correlations within the Ising chains (intra-chain correlations) are shown as open symbols and correlations between different Ising chains (inter-chain correlations) are shown as closed symbols.
	(b) Temperature evolution of the ferromagnetic intra-chain correlation length $\xi$ (colours as above).
	The solid lines show the exact expression [Eq.~(\ref{ising_correl}) in the text] for independent ferromagnetic Ising chains for $J_1|\mathbf{S}|^2=22.9$~K (red lines) and $J_1|\mathbf{S}|^2=36.8$~K (blue lines).}
\end{center}
\end{figure}
The radial spin correlation function $\langle S(0)S(r) \rangle$ obtained from RMC refinement of the $T=25$~K data is shown in Fig.~\ref{fig4_correlation}(a).
Correlations along the Ising spin chains  $\langle S(0)S(r_z) \rangle$ are FM and decay with distance $r_z$ along the chains.
This distance dependence can be fitted by an exponential decay, $\langle S(0)S(r_z) \rangle = \exp(-r_{z}/\xi)$, in agreement with the theoretical prediction for independent FM Ising chains.\cite{Baxter_1982}
By contrast, correlations between different chains are AFM, of smaller magnitude than the intra-chain correlations, and are not well described by an exponential decay; we consider them in more detail below.
The temperature dependence of the FM correlation length $\xi$ is shown in Fig.~\ref{fig4_correlation}(b). 
The values shown were determined by fitting $\langle S(0)S(r_z) \rangle$ with an exponential for each of the six measured temperatures. The fitting range was $0< r\,(\mathrm{\AA})<30$, except at $T=100$~K where the upper limit was reduced to $r=20~\mathrm{\AA}$ due to the very rapid decay of the correlations; in all cases, the quality of the fit was similar to that obtained for $T=25$~K. 
The value of $\xi$ decreases with increasing temperature, but is not negligible even at 100~K. 
Here we note a similarity with the theoretical expression,\cite{Baxter_1982} 
\begin{equation}
\xi=\frac{c}{2\ln\left[\coth\left(J_{1}|\mathbf{S}|^2/T\right)\right]},
\label{ising_correl}
\end{equation}
where $c=10.367$~\AA, which also has a ``long tail" at high~$T$.
Our observation of significant FM correlations well above \Tn\ is consistent with $\mu \rm{SR}$ data\cite{Takeshita_2006} and M\"{o}ssbauer studies on $^{159}$Eu-doped \cco,\cite{Paulose_2008} as well as with recent inelastic neutron scattering (INS) measurements, which report that well-defined magnetic excitations persist to $T\approx 150$~K.\cite{Jain_2013}

We note two qualifications regarding the RMC refinements. First, since RMC refinement is a stochastic process, the values of $\xi$ obtained from RMC represent lower bounds on the true values. Second, we found that making small changes to refinement parameters (in particular the intensity scale) resulted in significant variations in the absolute values obtained for the FM correlation length. This suggests that the data do not provide a strong constraint on the strength of the FM correlations. In order to estimate the uncertainties shown in Fig.~\ref{fig4_correlation} for $\langle S(0)S(r) \rangle$ and $\xi$, we considered the effect of changing the scale by a small amount from its refined value: $s \rightarrow s \pm \delta s$, where $\delta s =0.1s$ was chosen as the range over which it was possible to obtain reasonable fits to the data.

\begin{figure}[tb]
\begin{center}
\includegraphics[width=0.8\columnwidth]{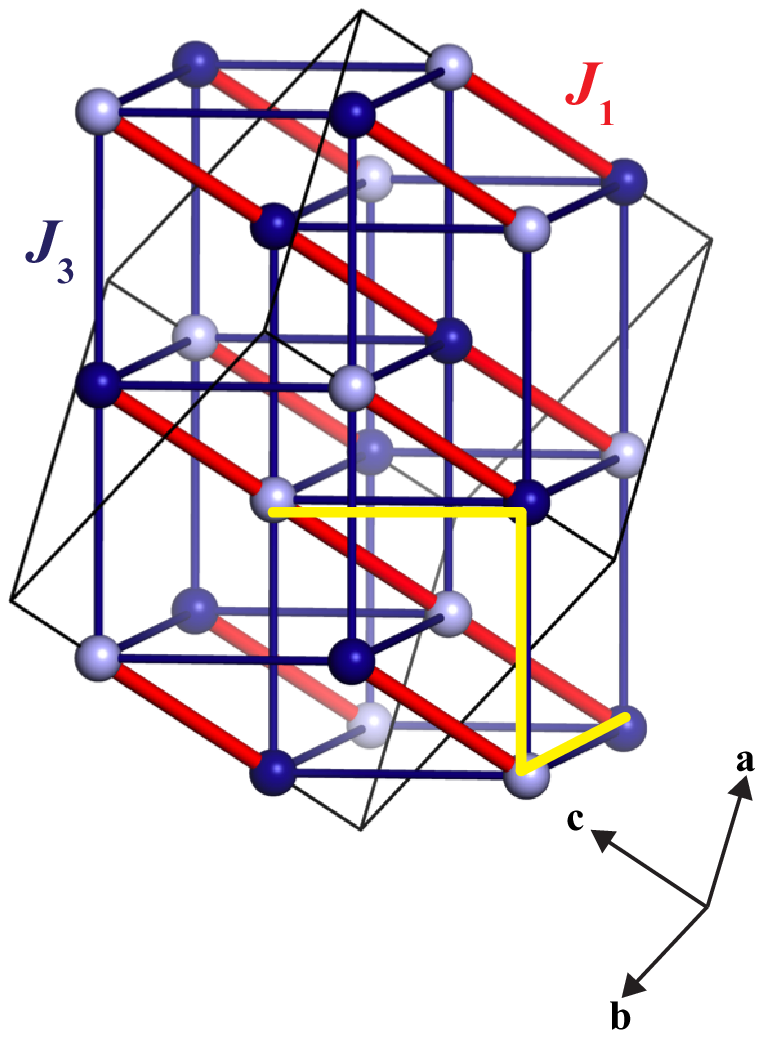}
\caption{\label{fig5_pathways}
	Magnetic interaction pathways in \cco\ (only Co$^{3+}_{\rm{II}}, S=2$ are shown).
	Ferromagnetic intra-chain couplings $J_1$ are shown as red lines and \AF ic (AFM) inter-chain couplings $J_3$ are shown as blue lines.
	Light and dark blue circles indicate the long-range magnetic ordering which would be obtained in the hypothetical case that $J_1=0$ with AFM $J_3$.
	Yellow lines show the shortest path between two sites within the same Ising chain if only $J_3$ pathways are used.}
\end{center}
\end{figure}
We now consider the extent to which the paramagnetic diffraction data can be modelled using a magnetic Hamiltonian.
Previous theoretical models have assumed that the spin chains behave like single Ising spins which are coupled by AFM interactions on a triangular lattice.\cite{Kudasov_2006,Kudasov_2007,Yao_2006}
However, these simplified models fail to reproduce the SDW phase below \Tn, so we consider the influence of the actual crystal structure on the magnetic interactions in our analysis.
Following previous work,\cite{Chapon_2009} we consider a model of Ising spins coupled by interactions between nearest neighbours ($J_1$ at 5.18~\AA), next-nearest neighbours ($J_2$ at 5.51~\AA) and third neighbours ($J_3$ at 6.23~\AA).
Our initial calculations revealed that our $T = 25$~K data can be well described by at least two different sets of parameters: either $J_1$ and $J_3$ (with $J_2=0$), or $J_1$ and $J_2$ (with $J_3=0$).
An independent determination of both $J_2$ to $J_3$ is therefore not possible based on only the diffuse scattering data.
However, it has been argued that $J_3$ should be of significantly greater magnitude than $J_2$, based on the shorter O--O bond distance for the $J_3$ pathway\cite{Fresard_2004} and on spin-dimer calculations.\cite{Chapon_2009} In order to proceed, we therefore assume initially that $J_{2}$ is negligible compared to $J_3$, and attempt to determine the values of only $J_1$ and $J_3$.
These interaction pathways are shown in Fig.~\ref{fig5_pathways}.
To this end, we performed direct Monte Carlo (DMC) simulations for different values of $J_3/J_1$ in the Ising Hamiltonian [Eq.~(\ref{ising_hamiltonian})], as described in Section~\ref{sec:technical}.
For each ratio of $J_3/J_1$ that we considered, we determined the ratio $T/J_1|\mathbf{S}|^2$ for which the calculated $I(q)$ most closely matched the $T=25$~K experimental data.
This procedure fixes the absolute value of $J_1$, and hence allows model and experimental data to be compared for each of the measured temperatures. 
 
Results of our simulations are shown in Figs.~\ref{fig3_IQ} and~\ref{fig4_correlation}.
We present results for $J_3/J_1=-0.02$, which is the ratio suggested in Ref.~\onlinecite{Chapon_2009}, and for $J_3/J_1=-0.10$, for comparison with our analysis in Section~\ref{sec:bulk}.
We also investigated a model with $J_3/J_1=-0.10$  and finite single-ion isotropy $D/J_1=32$, but the results described in this Section were very similar to those obtained for the Ising model ($D=\infty$), and so are not given here.
At most temperatures, both sets of interactions show good agreement with the data. There is a slight tendency for all the model calculations to lie below the data at low-$q$ and above the data at high-$q$, which may indicate a small inaccuracy in the assumed magnetic form factor.
The $J_3/J_1=-0.02$ model appears to fit the data better at $T=25$~K, but the $J_3/J_1=-0.10$ model is more successful at reproducing the shape of the main peak at $T=35$~K [Fig.~\ref{fig3_IQ}].
Both models are qualitatively consistent with the RMC results, in the sense that they show exponentially-decaying intra-chain FM correlations with a non-negligible value of $\xi$ at $T=100$~K, as well as weaker AFM inter-chain correlations [Fig.~\ref{fig4_correlation}].
The largest difference in $I(q)$ between the $J_3/J_1=-0.02$ and $J_3/J_1=-0.10$  models occurs in the region $q\lesssim 0.5$~\AA$^{-1}$, with smaller values of $|J_3/J_1|$ producing increased intensity as $q \rightarrow 0$.
Fig.~\ref{fig4_correlation}(b) shows that this increased intensity at low-$q$ is associated with a larger value of the FM correlation length $\xi$. 
Unfortunately, since the low-$q$ region is not accessed experimentally, the diffraction data are relatively insensitive to the values of $\xi$ and $J_3/J_1$.
In order to place a stronger restriction on $J_3/J_1$, it is necessary to consider other experimental evidence together with diffraction -- an approach we will follow in Section~\ref{sec:bulk}.
Nevertheless, a conclusion we can draw from Fig.~\ref{fig4_correlation}(b) is that, for small $|J_3/J_1|\lesssim 0.10$, the values of $\xi$ from DMC simulation agree with the exact result for \emph{independent} Ising chains [Eq.~(\ref{ising_correl})] with no fitting parameters.
Therefore, above \Tn\/, the functional form of the FM correlations is only weakly perturbed by a small $J_3$ interaction, even though \Tn\ itself increases with increasing $|J_3/J_1|$.
This result suggests that the coupling of $\rm{Co^{3+}}$ ions in the same chain by several $J_3$ pathways has negligible impact on the paramagnetic intra-chain correlations [Fig.~\ref{fig5_pathways}].
Clearly, this can only apply for small $J_3$; for $|J_3| >J_1$, the system is neither frustrated nor low-dimensional, since the $J_3$ interactions form a two-sublattice arrangement which would allow for the simple long-range ordering shown in Fig.~\ref{fig5_pathways}.

\begin{figure}[tb]
\begin{center}
\includegraphics[width=0.9\columnwidth]{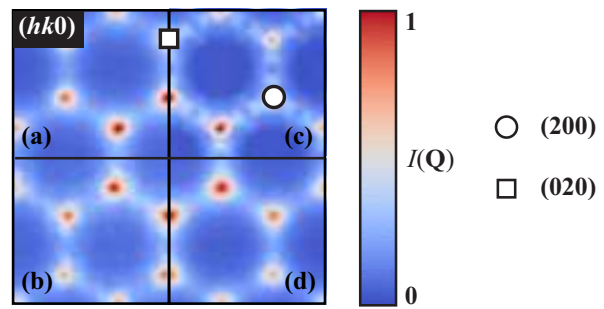}
\caption{\label{fig6_IvsQ_crystal}
	Calculated single-crystal diffuse scattering patterns at $T=25$~K for models of \cco\ discussed in the text:
	(a) direct Monte Carlo, $J_3/J_1=-0.10$ ($J_1|\mathbf{S}|^2=22.9$~K);
	(b) direct Monte Carlo, $J_3/J_1=-0.02$ ($J_1|\mathbf{S}|^2=36.8$~K);
	(c) reverse Monte Carlo, from fitting $T=25$~K powder data;
	(d) Wannier model of Ising spins on the triangular lattice calculated using Monte Carlo simulation at $T=1.5|J|$, where $J$ is the nearest-neighbour AFM interaction.  
	All panels show the $(hk0)$ reciprocal-space plane.
	The intensity scales are chosen so that the most intense feature in each panel is normalised to~1.}
\end{center}
\end{figure}
Before proceeding, we ask what further information on the paramagnetic spin correlations could be obtained from single-crystal diffraction data -- a question of topical interest given a recent report of a neutron scattering measurement on co-aligned crystals of \cco.\cite{Jain_2013} 
In Fig.~\ref{fig6_IvsQ_crystal}, we show the calculated $I({\bf{q}})$ in the $(hk0)$ reciprocal space plane for three models of \cco\ at $T=25$~K: the DMC model with $J_3/J_1=-0.10$ in~\ref{fig6_IvsQ_crystal}(a); the DMC model with $J_3/J_1=-0.02$ in~\ref{fig6_IvsQ_crystal}(b); and the RMC model obtained from fitting the powder data in~\ref{fig6_IvsQ_crystal}(c).
In each case the magnetic diffuse scattering pattern was calculated from the spin configurations using the general equation\cite{Squires_1978}
\begin{equation}
I(\mathbf{q})\propto [f(q)]^2\left|\sum_{i}\mathbf{S}_i^\perp\exp({\rm i}\mathbf{q}\cdot\mathbf{r}_i)\right|^2,
\end{equation}
where $f(q)$ is the $\rm{Co^{3+}}$ magnetic form factor,\cite{Brown_2004} $\mathbf{q}$ is the scattering vector, and $\mathbf{S}_i^\perp$ is the spin vector located at $\mathbf{r}_i$, projected perpendicular to $\bf{q}$. 
In all cases, the diffuse scattering takes the form of triangle-shaped peaks at the corners of the second Brillouin zone. 
The fact that similar results are obtained from both RMC refinement and a physically-sensible set of magnetic interactions suggests that our prediction of $I(\bf{q})$ is robust.
Our calculations also indicate that a single-crystal neutron-scattering measurement of the $(hk0)$ plane would provide rather little information on the relative strength of FM and AFM interactions. Instead, it would probably be necessary to measure one of the diffuse peaks along $\bf{c^*}$ in order to obtain the FM correlation length as a function of temperature.
Finally, in panel~\ref{fig6_IvsQ_crystal}(d) we show the calculated $I({\bf{q}})$ for the nearest-neighbour Ising \AF\ on the triangular lattice, a model due to Wannier\cite{Wannier_1950} which has previously been applied in several theoretical studies of \cco\ (see, \emph{e.g.}, Refs.~\onlinecite{Kudasov_2006,Kudasov_2007,Yao_2006}).
We find close agreement between panels~\ref{fig6_IvsQ_crystal}(a)--\ref{fig6_IvsQ_crystal}(c) and the Wannier model.
Indeed, the extent of this agreement is rather surprising, since the modulation of the diffuse scattering is sensitive to small changes in the magnetic interactions.\cite{Welberry_2011}
While the Wannier model clearly provides no information on the intra-chain correlations, it seems to provide a good qualitative description of the inter-chain spin correlations.
Taken as a whole, our results suggest a physical picture of paramagnetic \cco\ in which weak inter-chain correlations resemble the Wannier model, and the length-scale $\xi$ over which these correlations remain coherent along the $c$ axis can be calculated as if the chains behave independently. In this sense, above \Tn, the correlations along $c$ are effectively decoupled from those in the $ab$ plane.

\section{Bulk Properties}{\label{sec:bulk}}
\begin{figure}[tb]
\begin{center}
\includegraphics[width=0.85\columnwidth]{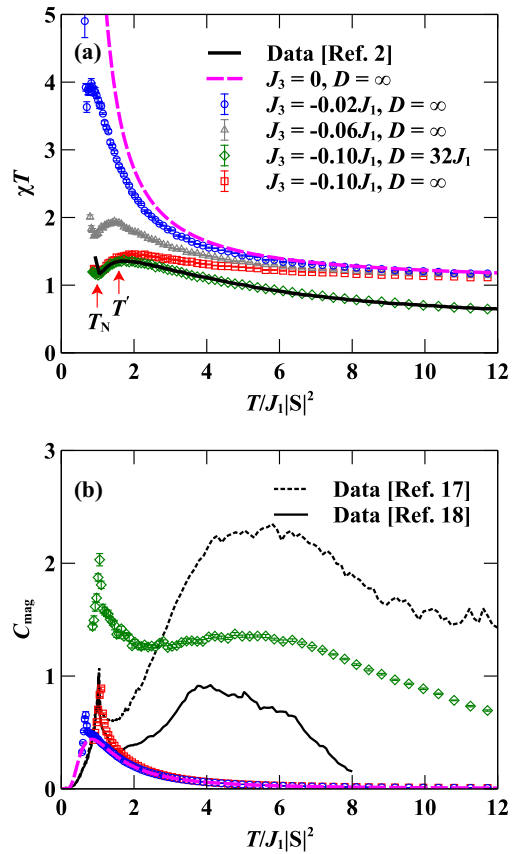}
\caption{\label{fig7_chi_C}
	(a) Magnetic susceptibility per spin of \cco\ parallel to the Ising chains.
	Experimental data (from Ref.~\onlinecite{Maignan_2004}) are shown as a solid black line.
	Results from direct Monte Carlo (DMC) simulations are shown for the Ising model with $J_3/J_1=-0.02$ (blue circles); $J_3/J_1=-0.06$ (grey triangles); $J_3/J_1=-0.10$ (red squares), and for the model with finite anisotropy $D/J_1=32$ and $J_3/J_1=-0.10$ (green diamonds).
	The exact result\cite{Baxter_1982} for FM $J_1$ only, $\chi T=\exp\left(2J_1|\mathbf{S}|^2/T\right)$, is shown for comparison (dashed magenta line).
	Better agreement with the data is obtained for the anisotropic model than for any of the Ising models.
	The experimental data have been scaled horizontally by matching the position of \Tn\ to the calculated curve with $J_3/J_1=-0.10$, and have been scaled vertically by a constant factor.
	(b) Magnetic heat capacity per spin of \cco.
	Data from Ref.~\onlinecite{Hardy_2003} and Ref.~\onlinecite{Hardy_2003a} are shown as dashed and solid black lines, respectively.
	Results from DMC simulations have the same symbols as in (a).
	The exact result\cite{Baxter_1982} for FM $J_1$ only, $C_{\mathrm{mag}}=\left(J_{1}|\mathbf{S}|^2/T\right)^2/\cosh^2\left(J_{1}|\mathbf{S}|^2/T\right)$, is shown as a dashed magenta line.}
\end{center}
\end{figure}
In this section we employ direct Monte Carlo simulations to calculate bulk properties of \cco\ in the paramagnetic phase.
Our motivation for considering bulk properties is the strong dependence of the limiting value $I(q\rightarrow0)$ on the ratio $J_3/J_1$, which suggests that the bulk susceptibility is sensitive to the strength of both FM and AFM interactions.
Here, Monte Carlo simulation has two important advantages over traditional approaches to fitting the susceptibility:\cite{Kageyama_1997,Martinez_2001,Maignan_2004,Hardy_2007} first, it is not limited by the number or type of interactions which can be modelled; second, it is not restricted to the high-temperature limit, remaining accurate as \Tn\ is approached from above.

In Fig.~\ref{fig7_chi_C}(a) the product of the magnetic susceptibility per spin and temperature, $\chi T$, is shown for different values of $J_3/J_1$.
The susceptibility is calculated parallel to the direction of the chains (\emph{i.e.}, along $c$), as described in Section~\ref{sec:technical}.
The results show two main features. First, there is an anomaly at the N\'{e}el temperature, $T_{\rm{N}}\sim J_1|\mathbf{S}|^2$.
The value of \Tn\ decreases with decreasing $|J_3/J_1|$, as one expects, since $T_{\rm{N}}=0$ in the limit of independent chains.\cite{Baxter_1982}
(We note that a value $T_{\rm{N}}=0$ cannot be obtained in Monte Carlo simulation due to the finite simulation size; for a model with only FM $J_1$ interactions, $\xi$ approaches the size of our simulation at $T\approx 0.4J_1|\mathbf{S}|^2$, \emph{i.e.}, significantly below the transition temperatures we observe for the $J_1$-$J_3$ model.)
Second, there is a broad peak at $T^\prime\gtrsim T_{\rm{N}}$, which sharpens and moves closer to \Tn\ as $|J_3/J_1|$ is decreased.
In this respect, our simulations reproduce experimental measurements of the parallel susceptibility,\cite{Maignan_2004,Cheng_2009} which also show a broad peak at $T^\prime\approx 1.6 T_{\rm{N}}=40$~K.
From Fig.~\ref{fig7_chi_C}(a), the experimental ratio of $T^\prime/T_{\rm{N}}$ implies that $0.06\lesssim |J_3/J_1|\lesssim 0.10$. 
Our results suggest that the $T^\prime$ peak arises from the development of AFM inter-chain correlations, since the diffraction measurements show the development of an AFM peak in $I(q)$ at $T\approx T^\prime$ [Fig.~\ref{fig3_IQ}], while the intra-chain correlations remain FM for all $T\leq100$~K [Fig.~\ref{fig4_correlation}(b)].

Although the Ising-model simulation with $J_3/J_1=-0.10$ shows qualitative agreement with the experimental data for $T\gtrsim T_{\rm{N}}$, it proved impossible to obtain a satisfactory fit of an Ising model over the whole temperature range: the problem is that the gradient of the Ising curve is too shallow at higher temperatures [Fig.~\ref{fig7_chi_C}(a)].
To explain this observation, it is necessary to relax our previous assumption that spins behave as purely Ising variables.
In general, an Ising approximation is valid for $T$ much smaller than the energy scale of magnetic anisotropy $D|\mathbf{S}|^2$ in the anisotropic Heisenberg Hamiltonian [Eq.~(\ref{heisenberg_hamiltonian})]. Literature values of $D$ range between $25\,\mathrm{K}\leq D\leq230\,\mathrm{K}$\cite{Kageyama_1997,Martinez_2001,Hardy_2007}, suggesting that the effect of finite anisotropy may not be negligible for all $T_{\rm{N}}< T\leq 300\,\mathrm{K}$.
Since $\chi T$ is measured experimentally by applying a magnetic field along the $c$ axis, spin fluctuations away from this axis for finite $D$ will cause the experimental $\chi T$ to be smaller than the Ising-model calculation.
To quantify the combined effect of $J_3$ and $D$, we performed DMC simulations of Eq.~(\ref{heisenberg_hamiltonian}) within the range $0.07\leq |J_3/J_1|\leq 0.14$ and $20\leq D/J_1\leq 40$.
For each set of interaction parameters, the calculated curve was fitted to the experimental data for $T>T_{\rm{N}}$ by first normalising both data and model horizontally by the position of \Tn\ and then calculating the vertical scale factor which minimised the sum of the squared residuals.
The best-fit parameters are $J_3/J_1=-0.10\pm 0.01$ and $D/J_1=32\pm 3$ (uncertainties represent the smallest interval in our grid).
The success of this procedure relies on the fact that $J_3$ and $D$ are only weakly correlated, since the effect of $J_3$ is most important for $T\gtrsim T_{\rm{N}}$, whereas the effect of finite $D$ is most pronounced at high $T$.
The corresponding fit is shown in Fig.~\ref{fig7_chi_C}(a), representing essentially quantitative agreement with the data throughout the paramagnetic regime. 

We can place our interaction parameters on an absolute scale using the location of $T_{\rm{N}}= 1.04J|\mathbf{S}|^2$.
Taking $|\mathbf{S}|^2=S(S+1)$ with $S=2$, we obtain $J_1= (4.0\pm 0.2)$~K, $J_3= (-0.40\pm 0.03)$~K, and $D= (130 \pm 20)$~K.
Our estimate of $J_1$ is significantly smaller than the  $J_1=13$~K obtained in Ref.~\onlinecite{Maignan_2004}, where the effect of finite $D$ was not considered when fitting $\chi T$ in the region $150 \leq T \leq 300$~K.
However, there is better agreement between our results and those of Ref.~\onlinecite{Hardy_2007}, in which the susceptibility was fitted to a model including both $J_1$ and $D$.
The most informative comparison is with recently-reported INS measurements,\cite{Jain_2013} which show a weakly-dispersive spin wave propagating along $\mathbf{c}^{*}$ with a large spin gap $\omega_0=27$~meV ($=310$~K).
The dispersion was fitted using linear spin-wave theory to obtain $J_1=4.9$~K and $D=\omega_0/2S=79$~K.
Taking into account higher-order corrections to the spin-wave theory leads to a modified equation for the spin gap, $D=\omega_{0}/(2S-1)$,\cite{Junger_2005} which yields $D=105$~K for the data of Ref.~\onlinecite{Jain_2013}.
With this correction, there is reasonable quantitative agreement ($\sim$20\%) between our results and INS measurements. We note that the orbital contribution to the magnetic moment has not been included in estimates of the interaction parameters, even though it represents a significant fraction of the total moment in \cco.
We can estimate the effective magnetic moment $\mu_{\mathrm{eff}}$ from the vertical scale factor using the relation  $\left(\chi T\right)_{\mathrm{expt}}=0.375(\mu_{\mathrm{eff}}/\mu_{\mathrm{B}})^2\left(\chi T\right)_{\mathrm{MC}}$, where $\left(\chi T\right)_{\mathrm{expt}}$ is given in $\mathrm{K\, emu\, mol^{-1}}$.
 We obtain $\mu_{\mathrm{eff}}=5.3\mu_{\mathrm{B}}$, which is consistent with theoretical studies ($\sim$5.7$\mu_{\mathrm{B}}$),\cite{Wu_2005} magnetisation measurements ($\sim$4.8$\mu_{\mathrm{B}}$),\cite{Maignan_2004} and an x-ray circular dichroism study ($\sim$5.3$\mu_{\mathrm{B}}$).\cite{Burnus_2006} 
 
Since considering finite $D$ introduces an additional parameter compared to an Ising model, it is important to ask whether it explains any measurements beyond the susceptibility.
One such measurement is the magnetic heat capacity, $C_{\rm{mag}}$.
In addition to the expected peak at \Tn, the experimental $C_{\rm{mag}}$ shows a broad high-$T$ peak centred at $T(C_{\rm{mag}})\approx120$~K.\cite{Hardy_2003,Hardy_2003a}
Although the presence of this peak is not in doubt, its intensity is quite uncertain due to the effect of the lattice subtraction, as a comparison of data from Ref.~\onlinecite{Hardy_2003} and Ref.~\onlinecite{Hardy_2003a} shows [Fig.~\ref{fig7_chi_C}(b)].
In Ref.~\onlinecite{Hardy_2003a} the high-$T$ peak was ascribed to the development of short-range FM intra-chain correlations, which release all the entropy in a fully one-dimensional model.
If this explanation is correct, then the large separation between $T(C_{\rm{mag}})$ and \Tn\ would suggest that \cco\ has much more one-dimensional character than the canonical Ising spin-chain compound $\rm{CoCl_2\cdot 2NC_5H_5}$, for which the high-$T$ peak is nearly obscured by the $T_{\rm{N}}$ peak.\cite{Takeda_1971} 
In Fig.~\ref{fig7_chi_C}(b) we plot the magnetic heat capacity from DMC simulations of the Ising model for $J_3/J_1=-0.10$ and $J_3/J_1=-0.02$.
For $J_3/J_1=-0.10$, the broad peak at $T(C_{\rm{mag}})$ is obscured by the sharp \Tn\ peak, and even for $J_3/J_1=-0.02$ it is only just visible as a shoulder on the \Tn\ peak.
Therefore an Ising model can only explain the existence of the high-$T$ peak if $|J_3/J_1|\ll0.02$.
We now consider our anisotropic Heisenberg model with $J_3/J_1=-0.10$ and $D/J_1=32$.
This model does show a broad peak at $T\approx 5J_1|\mathbf{S}|^2\approx 0.15D|\mathbf{S}|^2\approx120$~K, in agreement with the experimental data.
Our results are supported by quantum calculations of the anisotropic Heisenberg chain, which also predict an extra peak in $C_{\rm{mag}}$ at $T\approx 0.2DS(S+1)$ for large but finite $D$.\cite{Junger_2005}
However, an obvious problem is that our anisotropic Heisenberg curve does not collapse onto the Ising one at low temperature: this unphysical result is due to the failure of the classical approximation implied in Eq.~(\ref{heisenberg_hamiltonian}), which permits magnetic excitations which cost an infinitesimal amount of energy, resulting in a non-zero value for $C_{\rm{mag}}$ in the low-$T$ limit.\cite{Fisher_1964}
Hence the anisotropic Heisenberg curve should be compared with the data at high $T$, while the Ising curve provides a better description at low $T$. 

Finally, we consider our results in the context of the magnetic ordering identified by neutron diffraction. The first ordered state that develops at \Tn\ is a SDW with an incommensurate magnetic propagation vector $\mathbf{k}\approx \left(0,0,1.01\right)$.\cite{Agrestini_2008,Agrestini_2008a}
If it is assumed that $J_2$ is negligibly small, this propagation vector would require that $J_3/J_1= -0.02$.\cite{Chapon_2009}
The ratio $J_3/J_1= -0.10$ we have determined would produce a  shorter-wavelength modulation with $\mathbf{k}= \left(0,0,1.05\right)$.
However, as Fig.~\ref{fig7_chi_C}(a) shows, choosing $J_3/J_1= -0.02$ to agree with Ref.~\onlinecite{Chapon_2009} results in a much less successful description of the experimental susceptibility data.
These results may be reconciled if $J_2$ is not actually negligible compared to $J_3$.
Our simulations of the paramagnetic phase are not very sensitive to whether the AFM coupling is $J_2$ or $J_3$, or a combination of both; hence our value of $J_3$ can probably be interpreted as $J_{\rm{AFM}}$, where $J_{\rm{AFM}}\approx J_2+J_3$.
By choosing $J_3$ and $J_2$ appropriately, both the overall energy scale of AFM interactions and the experimental propagation vector can be reproduced. For $J_{\rm{AFM}}=-0.10J_1$, this occurs when  $J_2=J_3=-0.05J_1$.
We have checked that these values indeed yield a $\chi T$ curve which is qualitatively identical to the one shown previously for $J_3/J_1=-0.10$ ($J_2=0$).
In this context, we note that recent NMR results also indicate that $J_2$ may not be negligible.\cite{Allodi_2013}

\section{Conclusions}
In summary, polarised neutron diffraction measurements performed on \cco\ give clear evidence of the coexistence of the two different magnetic structures in this compound at low temperature: a spin density wave structure and a commensurate \AF ic structure.
The volume fraction of the phases is dependent on sample history in general, and on cooling rates in particular, with slow cooling rates providing easier access to the commensurate \AF ic phase.

Above \Tn, we performed reverse Monte Carlo refinements and direct Monte Carlo simulations to model the magnetic diffuse scattering data.
Our results show that intra-chain spin correlations remain ferromagnetic at all temperatures, decay approximately exponentially with distance along the chain, and exhibit a temperature-dependent correlation length which is well described by Ising's result for independent chains.
Inter-chain correlations are weaker and \AF ic, and have a diffraction pattern showing strong qualitative similarities to the canonical Wannier model.
Perhaps our most intriguing result is that important aspects of these two  models are simultaneously realised in paramagnetic \cco, \emph{i.e.}, intra- and inter-chain correlations are essentially decoupled.
The magnetic diffuse scattering data do not place strong constraints on the magnetic exchange constants, but by fitting published measurements of the single-crystal bulk susceptibility\cite{Maignan_2004} we obtain $J_1 \approx 4.0$~K, $J_3 \approx 0.40$~K, and $D\approx 130$~K.
In particular, we have shown that it is essential to include both \AF ic exchange ($J_2$ and/or $J_3$) and finite single-ion anisotropy $D$ in order to obtain a good description of bulk properties. 
The Hamiltonian we have described is the first to agree quantitatively with susceptibility data for all $T_{\rm{N}}< T\leq 300$~K; moreover, it is consistent with our neutron scattering results and correctly predicts the form of the magnetic heat capacity.\cite{Hardy_2003}

Given that not much is known about the behaviour of the CAFM phase in an applied field\cite{Agrestini_2011} and that for the SDW phase the magnetic states for a zero-field cooled sample and a sample that has previously been exposed to a high magnetic field are completely different on a microscopic level,\cite{Petrenko_2005,Fleck_2010} it would be extremely interesting to extend the neutron scattering measurements to an applied field.
Ideally such an experiment should be performed on an arrangement of single crystals with appreciable volume, which have recently been reported.\cite{Jain_2013}

\section{Acknowledgements}
We are grateful to M.~J.~Cliffe, L.~C.~Chapon, and P.~Manuel for valuable discussions.
J.~A.~M.~P. and A.~L.~G. gratefully acknowledge financial support from the STFC, EPSRC (EP/G004528/2) and ERC (Ref: 279705).

\bibliographystyle{Science}
\bibliography{prb_ca3co2o6_refs}
\end{document}